\documentclass{article}
\usepackage{graphicx}
\usepackage{amssymb}
\usepackage{times}
\usepackage{emulateapj}

\def\ref{\par\noindent\hang}

\def\spose#1{\hbox to 0pt{#1\hss}}
\let\approxlt=\lesssim
\let\approxgt=\gtrsim

\def\multleft#1{\hbox to size{\vbox {\halign {\lft{##}\cr #1}}\hfill}\par}
\def\multright#1{\hbox to size{\vbox {\halign {\rt{##}\cr #1}}\hfill}\par}

\def\boxit#1{\vbox{\hrule\hbox{\vrule\kern3pt\vbox{\kern3pt
          #1 \kern3pt}\kern3pt\vrule}\hrule}}

\def\cm{{\rm\thinspace cm}}

\def\erg{{\rm\thinspace erg}}

\def\K{{\rm\thinspace K}}
\def\keV{{\rm\thinspace keV}}
\def\km{{\rm\thinspace km}}

\def\Msun{\hbox{$\rm\thinspace M_{\odot}$}}

\def\s{{\rm\thinspace s}}

\def\pcmcu{\hbox{$\cm^{-3}\,$}}

\def\ergps{\hbox{$\erg\s^{-1}\,$}}

\def\kmps{\hbox{$\km\s^{-1}\,$}}

\def\ps{\hbox{$\s^{-1}\,$}}

\makeatletter
\@ifundefined{chapter}{\def\thebibliography#1{
\section*{References}
  \list
  {\relax}{\setlength{\labelsep}{0em}
        \setlength{\itemindent}{-\bibhang}
        \setlength{\itemsep}{\parskip}
        \setlength{\parsep}{0pt}
        \setlength{\leftmargin}{\bibhang}}
    \def\newblock{\hskip .11em plus .33em minus .07em}
    \sloppy\clubpenalty4000\widowpenalty4000
    \sfcode`\.=1000\relax}}%
{\def\thebibliography#1{
  \list
  {\relax}{\setlength{\labelsep}{0em}
        \setlength{\itemindent}{-\bibhang}
        \setlength{\itemsep}{\parskip}
        \setlength{\parsep}{0pt}
        \setlength{\leftmargin}{\bibhang}}
    \def\newblock{\hskip .11em plus .33em minus .07em}
    \sloppy\clubpenalty4000\widowpenalty4000
    \sfcode`\.=1000\relax}}
 
\newlength{\bibhang}
\let\@internalcite\cite
\def\cite{\@ifstar{\citey}{\citefull}}
\def\citefull{\def\astroncite##1##2{##1\ ##2}\@internalcite}
\def\citey{\def\astroncite##1##2{##1\ (##2)}\@internalcite}
\def\citeyear{\def\astroncite##1##2{##2}\@internalcite}
\def\citename{\def\astroncite##1##2{##1}\@internalcite}
\def\@citex[#1]#2{\if@filesw\immediate\write\@auxout{\string\citation{#2}}\fi
  \def\@citea{}\@cite{\@for\@citeb:=#2\do
    {\@citea\def\@citea{; }\@ifundefined
       {b@\@citeb}{{\bf ??}\@warning
       {Citation `\@citeb' on page \thepage \space undefined}}%
{\csname b@\@citeb\endcsname}}}{#1}}
 
\def\@cite#1#2{#1\if@tempswa #2\fi} 
\def\@biblabel#1{}
 
\def\astroncite#1#2{#1\ #2}
\makeatother

\begin{document}

\title{On the inability of Comptonization to produce\\ the broad X-ray iron
lines observed in Seyfert nuclei}

\author{Christopher~S.~Reynolds\altaffilmark{1,2} and J\"orn
Wilms\altaffilmark{3}} 

\altaffiltext{1}{JILA, Campus Box 440, University of Colorado,
Boulder CO~80303}

\altaffiltext{2}{Hubble Fellow}

\altaffiltext{3}{Institut f\"ur Astronomie und
Astrophysik--Astronomie, University of T\"ubingen, Waldh\"auser
Stra\ss{}e 64, D-72076 T\"ubingen, Germany}

\begin{abstract}
It has recently been suggested that Compton downscattering may give rise to
the broad iron lines seen in the X-ray spectra of Seyfert 1 galaxies.  This
challenges the standard model in which these lines originate from the
innermost regions of the black hole accretion disk with Doppler shifts and
gravitational redshifts giving rise to the broadened line profile.  Here,
we apply observational constraints to the Compton downscattering model for
MCG$-$6-30-15 and NGC~3516, the two best cases to date of Seyfert galaxies
with relativistically broad lines.  We show that the continuum source in
MCG$-$6-30-15 required by the constrained model violates the black body
limit.  In the case of NGC~3516, only a very small region of parameter
space is compatible with the constraints.  Hence, we conclude that the
Comptonization model is not a viable one for the broad line seen in these
two objects.  The accretion disk model remains the best interpretation of
these data.
\end{abstract}


\section{Introduction}

The X-ray study of Seyfert nuclei and other types of active galactic nuclei
(AGN) has been energized for the past few years by the observation of
relativistically broad iron K$\alpha$ lines in their X-ray spectra (Tanaka
et al. 1995; Nandra et al. 1997; Reynolds 1997).  In particular, the
Seyfert galaxy MCG$-$6-30-15 has become an important testing ground for
models of broad iron line formation.  A long observation of MCG$-$6-30-15
by the {\it Advanced Satellite for Cosmology and Astrophysics (ASCA)}
revealed a high signal-to-noise broad iron line with a velocity width of
$\sim 10^5\kmps$ and a profile which is skewed to low energies (Tanaka et
al. 1995).  The excitement stirred by these studies is due to the widely
held belief that the iron lines originate from the surface layers of an
accretion disk which is in orbit about a supermassive black hole, and that
the line width and profile provide a direct probe of the velocity field and
strong gravitational field within a few Schwarzschild radii of the black
hole.  Models of line emission from the inner regions of a black hole
accretion disk (e.g., Fabian et al. 1989; Laor 1991) fit the observed line
profiles well.

The suggestion that we are observing the immediate environment of an
accreting supermassive black hole is a bold one and certainly warrants a
critical examination.  In this spirit, Fabian et al. (1995; hereafter F95)
examined a number of alternative hypotheses for the origin of these broad
iron lines including models in which the line is produced in an outflow or
jet, and models in which the line is intrinsically narrow (or even absent)
and a complex underlying continuum mimics the broad line.  Both of these
classes of models were found to be unphysical or did not reproduce the
observed spectrum.

Another alternative model, first proposed by Czerny, Zbyszewska
\& Raine (1991) but also considered by F95, is one in which the
iron line is intrinsically narrow (i.e., emitted in slowly moving
material which is very far from a compact object) and then
broadened to the observed profile by Compton downscattering in
matter that surrounds the source of line photons.  F95 rejected
this model on the basis that the Comptonizing cloud must have a
radius of $R<10^{14}\cm$ in order to maintain the required high
ionization state and that, with such a small radius,
gravitational effects from a central $10^7\Msun$ black hole would
be important anyway for determining the line profile.  The
principal aim of F95 was to demonstrate the need to include
strong gravity in any model of the iron line, so they terminated
their chain of reasoning at that point.  The question remains,
however, as to whether Compton downscattering has a significant
affect on the line profile or whether we can interpret iron line
observations in terms of naked accretion disk models.

Misra \& Kembhavi (1998) and Misra \& Sutaria (1999; hereafter
MS99) have recently developed the Comptonization model further.
In their current model, they suggest that a cloud with optical
depth $\tau=4$ and temperature $kT\approxlt 0.5\keV$ surrounds
the central engine.  The upper limit to the temperature of the
Compton cloud comes from the fact that the iron K$\alpha$ line
photons need to be primarily {\it downscattered}, rather than
upscattered, in order to reproduce the observed line profile.
The central engine produces the continuum emission which keeps
the cloud ionized, and a narrow iron line which is Compton
broadened to the observed width.  They show that the resultant
line profiles can be brought into good agreement with the {\it
ASCA} observations.

A direct prediction of the Comptonization model (F95, MS99) is that the
multiple Compton scatterings should produce a break in the spectrum of
the power-law continuum radiation at approximately $E_{\rm br}\sim
m_{\rm e}c^2/\tau^2$ (i.e., $\sim 30$--$40\keV$).  Recently, it has been
reported that {\it BeppoSAX} (Guainazzi et al. 1999) observations
constrain the location of the continuum break to be at energies greater
than $100\,\keV$ , thereby arguing against the Comptonization model
(Misra 1999).  However, a robust determination of the continuum break is
not completely straightforward since it depends upon the parameters
assumed for the Compton reflection component (e.g., see Lee et
al. 1999).  Thus, while the lack of a spectral break at 30--40\,keV
remains the most compelling argument against the Comptonization model,
it is interesting to consider constraints on the Comptonization model
that are independent of a continuum spectral break.

In this paper, we apply a number of observational constraints to the
MS99 model.  We focus on the case of the iron line in MCG$-$6-30-15,
but also address the line in NGC~3516, the other high signal-to-noise
case of a relativistically broad line.  We show that the continuum
source in MCG$-$6-30-15 required by the constrained model violates
thermodynamic limits (i.e., the ``black body'' limit).  We also show
that only a very small region of parameter space is open to the
Comptonization model in the case of NGC~3516.  Hence, we conclude that
the Compton downscattering model is not a viable model for the broad
iron lines in one, and possibly both, of these sources.

\section{Constraints from continuum variability in MCG$-$6-30-15}

The iron line in MCG$-$6-30-15 has been observed to change flux
and profile on timescales of $10^4\s$ (Iwasawa et al. 1996,
1999).  This is the shortest timescale on which detailed line
changes can currently be probed and there may indeed be line
variability on shorter timescales.  MS99 note that such
variability is consistent with the line originating from a
Compton cloud of size $R\sim 10^{14}\cm$.

However, in the Comptonization model, the continuum photons also
pass through the same Comptonizing medium as the iron line
photons.  Thus, continuum variability can be used to place much
tighter constraints on the size of the cloud.  Any variability of
the central source would be smeared out as the photons random
walk through the cloud on a timescale of
\begin{equation}
t_{\rm MS}\sim \frac{R\tau}{c}.
\end{equation}
Appreciable continuum variability in MCG$-$6-30-15 is observed on
timescales down to $t_{\rm obs}\sim 100\s$ (Reynolds et al. 1995;
Yaqoob et al. 1997).  Since we must have $t_{\rm obs}\approxgt
t_{\rm MS}$, an upper limit on the Compton cloud is $R_{\rm
cloud}= 10^{12}\cm$, two orders of magnitude less than the size
assumed in MS99.  Assuming a geometrically thick cloud and solar
abundances, the density of the material is $n_{\rm H}\approxgt
5\times 10^{12}\pcmcu$.

In assessing the robustness of this constraint, it should be noted that
the iron line in MCG--6-30-15 is always observed to be broad (although
the width of the line does indeed vary, e.g. Iwasawa et al. 1996), and
the source is always observed to vary its flux with a temporal power
spectrum that extends down to 100\,s timescales (Lee et al. 1999b; Nowak
\& Chiang 1999; Reynolds 1999).  Thus, it is difficult to support a
model in which the Compton cloud is sometimes present (producing a broad
line and a slowly varying continuum) and sometimes absent (producing a
narrow line and a rapidly varying continuum).

\section{The Compton temperature and the black-body limit}

\begin{figure*}
\includegraphics[width=0.45\textwidth]{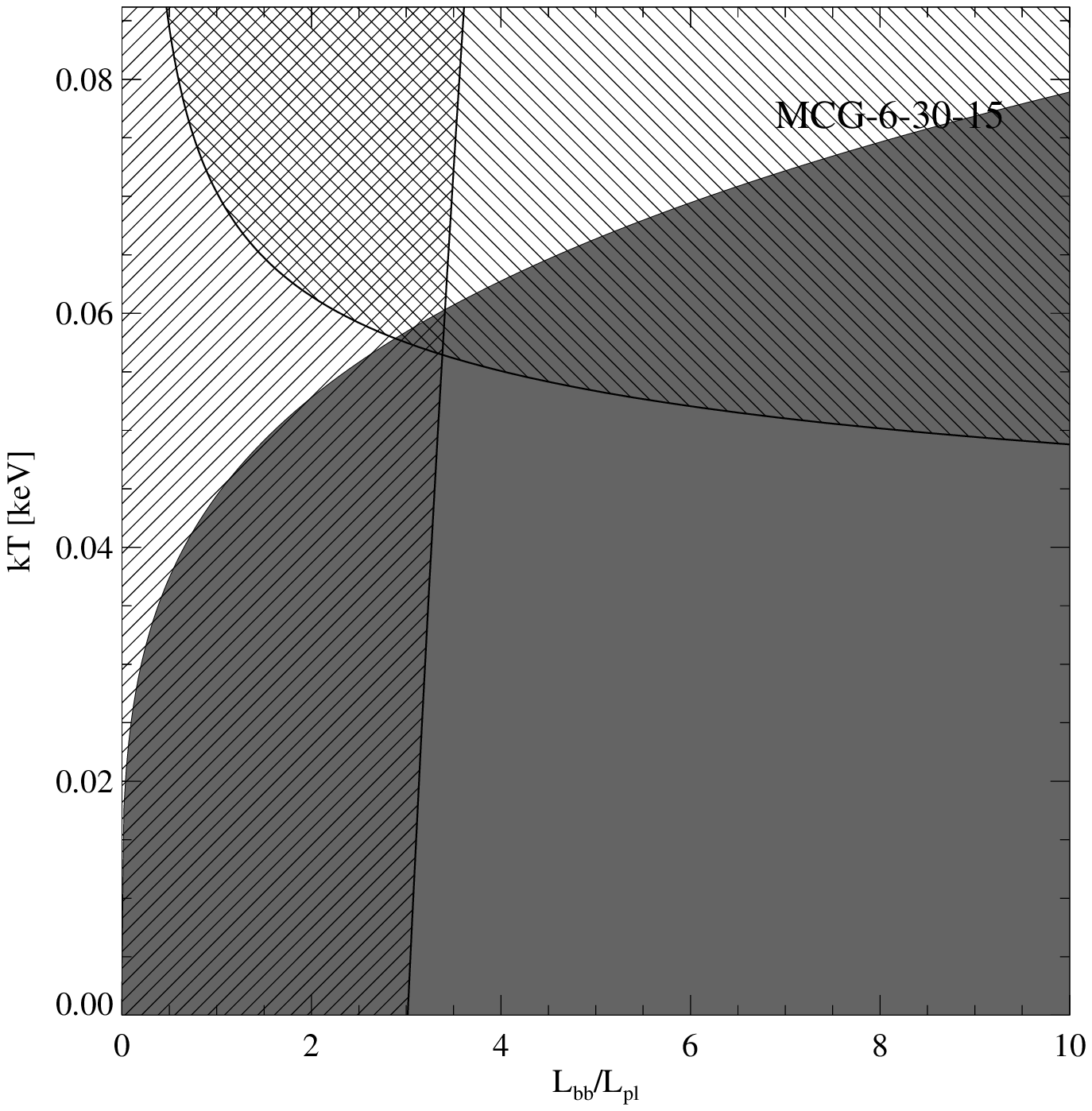}
\includegraphics[width=0.45\textwidth]{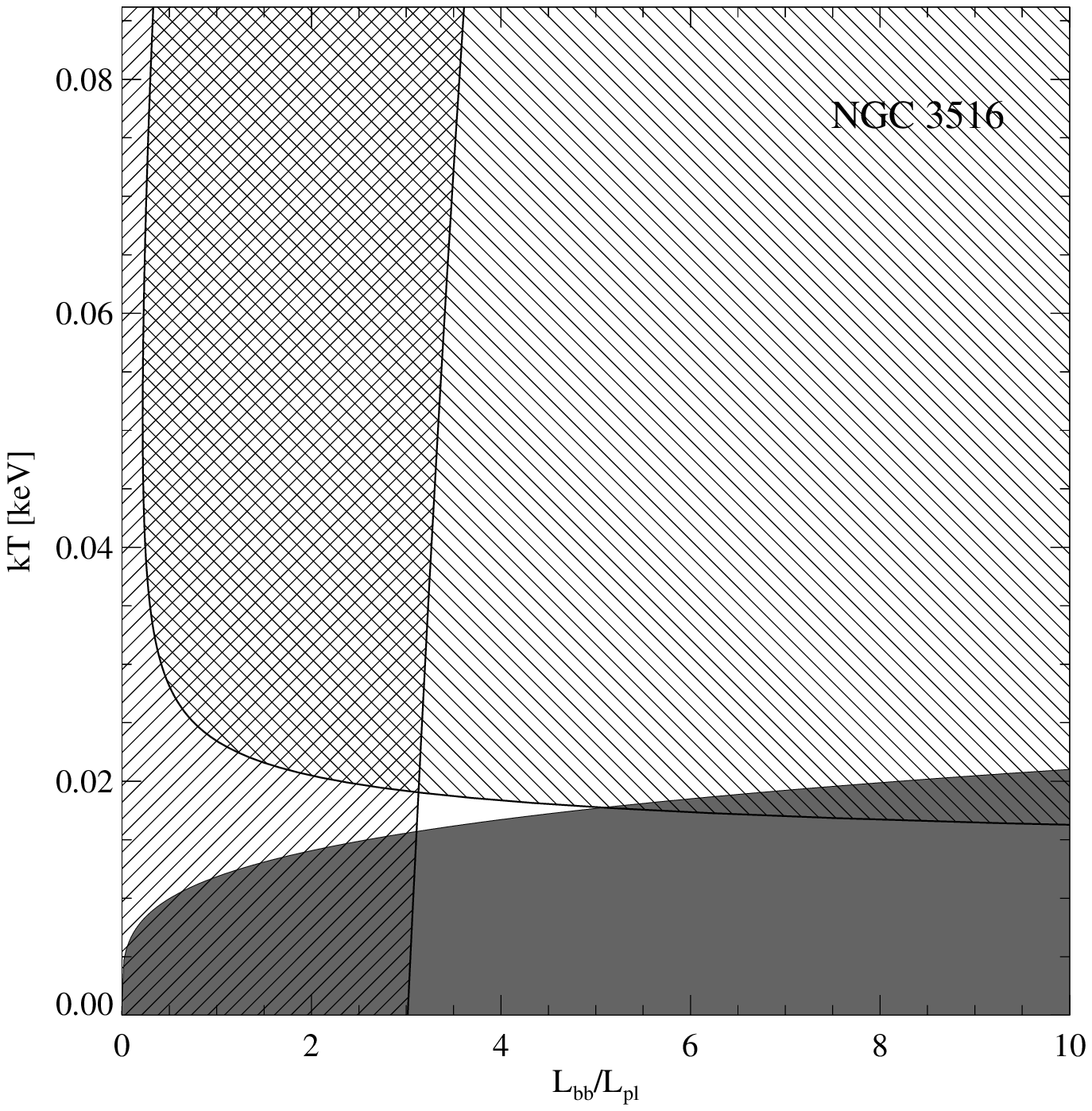}
\caption{Constraints diagrams for the Comptonization model
applied to MCG$-$6-30-15 (left) and NGC~3516 (right).  The almost
vertical line corresponds to a Compton temperature of $kT_{\rm
C}=0.5\keV$, with the region left of the line being excluded
since it would produce too much Compton upscattering of the iron
line photons.  The region shaded with lines of negative slope is
forbidden since it would produce a soft excess in the {\sl ASCA}
(MCG$-$6-30-15) or {\sl BeppoSAX} (NGC~3516) bands (which is not
observed).  The shaded region is forbidden since the source would
violate the black body limit.}
\end{figure*}

In the situation postulated by the MS99 model, the temperature of
the Compton cloud will be locked to the Compton temperature of
the (local) radiation field.  We model the continuum spectrum of
the central source as the superposition of a black body spectrum
(which may represent thermal emission from an accretion disk) and
a power-law spectrum with energy index $\alpha=1$ which extends
up to hard X-ray energies (which may be identified as accretion
disk photons that have been subjected to multiple Compton
upscattering by an accretion disk corona).

The flux at the inner edge of the Compton cloud is then given by
\begin{equation}
F_\nu=\frac{h \nu^3 L_{\rm bb}}
           {2\sigma_{\rm SB}T^4 c^2  R_{\rm in}^2(\exp(h\nu/kT)-1)} 
     +\frac{L_{\rm pl}f(\nu)}{4\pi R_{\rm in}^2\Xi},
\end{equation}
where $R_{\rm in}$ is the inner radius of the Compton cloud,
$L_{\rm bb}$ is the luminosity of the black body component,
$\sigma_{\rm SB}=5.67\times 10^{-5}\,\rm
erg\,cm^{-2}\,s^{-1}\,K^{-4}$ the Stefan-Boltzmann constant,
$L_{\rm pl}$ is the luminosity in the power-law component,
$f(\nu)=\nu^{-1}$ in the range $\nu_{\rm min}<\nu<\nu_{\rm max}$
(and zero elsewhere), and $\Xi$ is given by
\begin{equation}
\Xi=\ln\left(\frac{\nu_{\rm max}}{\nu_{\rm min}}\right).
\end{equation}
Guided by the hard X-ray observations of MCG$-$6-30-15 (e.g., Lee et
al. 1999), the parameters describing the power-law component are fixed
to have the following values:
\begin{eqnarray}
h\nu_{\rm min}&=&0.1\keV,\\
h\nu_{\rm max}&=&50\keV,\\
L_{\rm pl}&=&5\times 10^{43}\ergps
\end{eqnarray}
The resulting Compton temperature is given by
\begin{equation}
T_{\rm C}=\frac{1}{1+{\cal L}}\left( T\,{\cal
L}+\frac{h(\nu_{\rm max}-\nu_{\rm min})}{4k\Xi}\right),
\end{equation}
where ${\cal L}$ is the ratio of the black body luminosity to the
power-law luminosity:
\begin{equation}
{\cal L}=\frac{L_{\rm bb}}{L_{\rm pl}}.
\end{equation}
The line corresponding to a Compton temperature of $T_{\rm C}=0.5\keV$
on the $({\cal L},T)$-plane is shown on Fig.~1a, and the forbidden
region of parameter space (giving $T_{\rm C}>0.5\keV$) is shaded with
lines of positive gradient.

For completeness, it should be noted that the above expression
for the Compton temperature is only strictly valid due to the
soft nature of our spectrum.  The Compton temperature depends, of
course, on the form of the radiation field inside the cloud.
Ignoring downscattering, this field is greater than the external
radiation field by a factor of $\tau$.  For the high-energy
radiation ($h\nu\approxgt 50\keV$), $\tau$ has an energy
dependence due to Klein-Nishina corrections, thereby affecting
the Compton temperature.  The neglect of
downscattering is also invalid at these energies.  However, these
corrections to the Compton temperature have a negligible effect
in our case.

The {\it ASCA} observation shows no evidence for a soft excess
component in MCG$-$6-30-15 across the entire well-calibrated
spectral range of the solid-state imaging spectrometers (SIS;
0.6--10\,keV).  Thus, we impose the condition that the black-body
flux at 0.6\,keV is less than the power-law flux at the same
energy:
\begin{equation}
\frac{h \nu^3 L_{\rm bb}}{2\sigma_{\rm SB}T^4 c^2R_{\rm
in}^2(\exp(h\nu/kT)-1)}<\frac{L_{\rm pl}f(\nu)}{4\pi R_{\rm in}^2\Xi}
\end{equation}
The region on the $({\cal L},T)$-plane forbidden by this constraint is
shaded with lines of negative gradient in Fig.~1a.

Finally, we make the observation that there is a fundamental limit to the
black body luminosity which is imposed by thermodynamics:
\begin{equation}
L_{\rm bb}<4\pi R_{\rm max}^2 \sigma_{\rm SB} T^4
\end{equation}
where $R_{\rm max}$ is the maximum allowed size of the black body source.
Since the continuum source is hypothesized to be interior to the Compton
cloud, we must have $R_{\rm max}\approxlt R_{\rm cloud}$.  The region of the
$({\cal L},T)$-plane forbidden by this constraint is shown in solid-shade
in Fig.~1a.

We see that applying these three constraints eliminates all
regions of the $({\cal L},T)$-plane.  One must conclude that the
Compton cloud model discussed by Misra \& Kembhavi (1998) and
MS99 is not valid in the case of MCG$-$6-30-15.

NGC~3516 also displays a strong broad iron line that has been observed at
high signal-to-noise with {\it ASCA} (Nandra et al. 1999).  We have also
examined constraints on the Comptonization model for this iron line.
Continuum variability in this object is observed on timescales down to
$\sim 2000\s$ (Edelson \& Nandra 1998; K. Nandra, private communication),
giving a maximum size of $R_{\rm cloud}\sim 2\times 10^{13}\cm$ for the
Comptonizing cloud, rather larger than the case of MCG$-$6-30-15.  Also,
{\it BeppoSAX} observations fail to see a soft excess in the X-ray spectrum
all of the way down to $0.2\keV$ (Stirpe et al. 1998).  Noting that $L_{\rm
pl}\approx 1\times 10^{44}\ergps$ produces the constraint diagram shown in
Fig.~1b.  It is seen that these constraints eliminate all but a very small
region of parameter space.  
Thus, although the broad line in NGC~3516 could in principle be
explained with the Comptonization model, the amount of fine
tuning necessary for finding the line parameters makes the model
improbable in this case.

\section{Discussion}

It should be stressed that we have used conservative parameters in our
assessment of these observational constraints.  In particular, we assume
that the power-law component of the continuum emission possesses an energy
index of $\alpha=1$ (corresponding to a photon index of $\Gamma=2$) and a
high energy cutoff of $50\keV$.  In fact, the overall X-ray spectrum is
harder than this (especially once the Compton reflection component is
accounted for) and the high energy cutoff may well occur at rather higher
energies.  Either of these effects will raise the Compton temperature of
the power-law component and require an even cooler black body component in
order to cool the Compton cloud below the $0.5\keV$ limit.  It should
also be noted that we have ignored any infra-red emission from the
continuum source.  Due to the high densities of the matter in the Compton
cloud, IR emissions redwards of $\sim 10\mu {\rm m}$ will be free-free
absorbed and act to {\it heat} the cloud rather than Compton cool it.
Again, the neglect of the IR emissions is a conservative assumption for our
purposes.

There is another, independent, problem faced by the Compton cloud model:
it is very difficult to maintain the required ionization state.  F95
treated this problem by considering the required cloud size necessary to
acheive some critical ionization parameter $\xi_{\rm c}\equiv L_{\rm
ion}/nR^2$. According to F95, the AGN spectrum of Mathews \& Ferland
(1987), $\xi_{\rm c}=10^4\erg\cm\ps$ can be considered the point at
which a photoionized plasma becomes completely ionized.  Using the
observed luminosity of MCG$-$6-30-15, they deduced that the cloud must
have a size $R<10^{14}\cm$ in order to achieve at least this critical
ionization parameter.  As we will now show, this is a very conservative
argument and, in fact, ionization balance imposes much more severe
limits on the cloud size.

While the formal ionization parameter may be very high, the very soft
continuum spectrum postulated by MS99 may still have trouble fully
ionizing the iron throughout the whole cloud.  To see this, note that
all continuum photons capable of ionizing hydrogen like iron (Fe\,{\sc
xxvi}) reside in the power law component of the continuum.  The continuum
source in MCG$-$6-30-15 emits {Fe\,\sc xxvi} ionizing photons at a
rate
\begin{equation}
N_{\rm ion}\approx\frac{L_{\rm pl}}{E_{\rm ion}\Xi},
\end{equation}
where $E_{\rm ion}=9.3\keV$ is the ionization potential of Fe\,{\sc
xxvi}.  This evaluates to $N_{\rm ion}\approx 3\times
10^{50}\ps$.  The radiative recombination rate of the postulated
Compton cloud, on the other hand, is given by
\begin{equation}
N_{\rm rec}\approx\frac{4\pi}{3}R^3 n^2 A_{\rm
rad}\left(\frac{T}{10^4\K}\right)^{-X_{\rm rad}}
\end{equation}
where the coefficients $A_{\rm rad}$ and $X_{\rm rad}$ are given by Shull
\& van Steenberg (1982).  For a temperature of $kT=0.5\keV$ and
$R=10^{12}\cm$, this gives $N_{\rm rec}=3\times 10^{50}\ps$.  Thus, there
are just enough ionizing photons present {\it in the entire power law tail} to
ionize the hydrogen-like iron.  If the temperature of the Compton cloud is
below 0.5\,keV, or the radius of the cloud is larger\footnote{Note that the
quantity $nR$ is proportional to the optical depth of the cloud and so is
fixed by the width of the broad iron line.}, it will be impossible to
photoionize the cloud.  Very large iron edges would then be present in the
observed X-ray spectrum, contrary to observations.
Thus, ionization balance imposes a size limit of $R\approxlt
10^{12}\cm$, independently of continuum variability constraints.

Finally, we address whether there are reasonable modifications that can be
made to the MS99 scenario that will avoid the constraints imposed in this
paper.  There are three such modifications that we should consider.
Firstly, if the geometry is such that the X-ray continuum source is viewed
directly (rather than through the Compton cloud), one might imagine that
the size of the Compton cloud and the X-ray continuum variability would be
decoupled thereby relaxing the constraints discussed above.  An example of
such a geometry is if the Compton cloud forms a torus around the central
X-ray source.  In such a geometry, the X-ray continuum source illuminates
and ionizes the observed face of the Compton cloud and powers iron line
fluorescence from an optical depth of $\tau\sim 4$ into the cloud.
However, in this case, one would expect ionized iron lines (from the
ionized zones that overlay the near-neutral zones in the Compton cloud)
rather than the observed cold iron lines.  Also, the illuminated surface of
the Compton cloud, which must be highly ionized so as not to be a strong
narrow iron line emitter, would act as a Compton mirror and smear out the
observed continuum variability, even though the continuum source is viewed
directly.
Of course, any such modification to the basic Comptonization model in
which the Compton cloud is allowed to be bigger than $R\sim 10^{12}\cm$
must be subject to the ionization problem described above.

Secondly, a large region of parameter space would open up if the Compton
cloud experienced a different soft continuum to that observed (i.e. if the
soft excess can be `hidden' from view).  Noting that the black body
component must scatter though the same parts of the Compton cloud that
broadens the iron line (in order to Compton cool it), one concludes that
the black body photons and broad iron lines photons will follow very
similar paths through the system.  Hence, it is impossible to hide the soft
excess emission from view in a system in which we observe a Compton
broadened iron line.

Thirdly, the black body limit can be bypassed if the soft continuum source
is placed outside of the Compton cloud.  While it is difficult to construct
rigorous arguments against this case, we consider that placing a powerful
($L_{\rm bb}/L_{\rm pl}>3$) soft continuum source at large distances from
the central hard X-ray continuum source is an ad-hoc solution.

\section{Conclusions}

In this work, we have constrained the Compton cloud model for the broad
iron line in both MCG$-$6-30-15 and NGC~3516 by considering two
observational constraints
which are independent of the detection of a spectral break in the
continuum spectrum:
the continuum variability timescale and the absence of an observed soft
excess.  We have then demonstrated that the constrained model requires a
continuum source which violates the black body limit.  We also point out
that the difficulty of photoionizing the Compton cloud to the required
levels.  Thus, we rule out the Comptonization model for the broad iron
line in MCG$-$6-30-15, and show that fine tuning is required in order
for the model to explain the line in NGC~3516.  We conclude that
the combination of relativistic Doppler shifts and gravitational
redshifts still provides the best explanation for the broad iron
lines seen in AGN. 

\acknowledgements

We are indebted to Jim Chiang, Andrew Fabian, Mike Nowak, and Firoza
Sutaria for insightful discussions throughout the course of this work.
We are also grateful to the anonymous referee for several useful
suggestions.  We thank the Aspen Center for Physics for their
hospitality during the X-ray Astrophysics Workshop in August 1999, at
which time this work was started.  CSR appreciates support from Hubble
Fellowship grant HF-01113.01-98A.  This grant was awarded by the Space
Telescope Institute, which is operated by the Association of
Universities for Research in Astronomy, Inc., for NASA under contract
NAS 5-26555.  We also appreciate support from NASA under LTSA grant
NAG5-6337 and the RXTE guest observer grant NAG5-7339 as well as
Deutsche Forschungsgemeinschaft grant Sta~173/22.

\end{document}